\def\kms{\hbox{km s$^{-1}$\,}}

\def\hii{H{\sc ii}\,}

\def\msun{M$_\odot$\,}
\def\mjyb{\hbox{mJy beam$^{-1}$}}
\def\msunyr{\hbox{M$_\odot$ yr$^{-1}$}}
\def\cm2{cm$^{-2}$}
\def\cm3{cm$^{-3}$}

\def\Ha{H$\alpha$}

\def\ha{H$\alpha$}

\def\gra{$^{\circ}$}
\def\arcmin{$'$}
\def\fdg{$'$.}

\documentclass[structabstract]{aa}

\usepackage{natbib}
\usepackage{graphicx}
\usepackage{txfonts}
\begin{document}
   \title{ NANTEN $^{12}$CO  (J=1$\rightarrow$0) observations around the star WR 55}

\author{N. U. Duronea\inst{1}
          \and
           E. M. Arnal\inst{1,2}
          \and J. C. Testori\inst{1}    
          }
\institute{Instituto Argentino de Radioastronomia, CONICET, CCT-La Plata, 
 C.C.5., 1894, Villa Elisa, Argentina   \email{duronea@iar.unlp.edu.ar}\and Facultad de Ciencias Astron\'omicas y Geof\'isicas, Universidad Nacional de La Plata,Paseo del Bosque s/n, 1900 La Plata,  Argentina} 

\date{Received 2011 December; accepted 2012 February}

  \abstract
   {A complete study of the molecular and ionized gas in the environs of the nebula RCW 78 around WR 55  is presented.  }
   {The aim of this work is to   investigate the spatial distribution, physical characteristics,  and  kinematical properties  of the molecular gas linked to the galactic nebula  RCW 78  to  achieve a better understanding of its interaction with the star and with the ionized gas.  
  }
   { This study was based on $^{12}$CO(1-0) fully sampled observations of a region of  $\sim$ 0\fdg45 in size around the star WR 55 and the nebula RCW 78 obtained with the 4-m NANTEN telescope, radio continuum archival  data at 1.4 and 4.85 GHz, obtained from SGPS  and PMNRAO Southern Radio Survey, respectively, and available infrared MIPSGAL images at 24 $\mu$m.}
   {A molecular gas    component in the velocity range from $\sim$ --58 to --45 \kms, compatible with the velocity of the ionized gas, was found to be associated with the optical nebula. Adopting a distance of $\sim$ 5 kpc, the mass of this molecular component is about 3.4 $\times$ 10$^4$ M$_{\odot}$. The analysis of the molecular data revealed the presence of a velocity gradient, in agreement with the H$\alpha$ line.

New radiocontinuum  flux density determinations confirm the thermal nature of RCW 78. This indicates that the ionized gas in RCW 78 arises from photoionization of the molecular gas component in the velocity range from $-$58 \kms to $-$45 \kms. 

A molecular concentration at a velocity of $-$56.1 \hbox{\kms}  (identified as C1) is very likely associated with the star HD 117797 and with a collection   of candidate YSOs, lying at a distance of 3.9 kpc, while the rest of the molecular gas at velocities between --56 and -46 \kms constitute   an incomplete ring-like structure which expands around WR 55   at a velocity of about $\sim$ 5 \kms. Mechanical energy and time requirements indicate that WR 55 is very capable of sustaining the expansion of the nebula.
}
   {}

   \keywords{ISM: molecules, radio continuum, ISM: H{\sc ii} regions, Individial object: RCW 78, Stars: WR 55, HD 117797}
   \maketitle

%________________________________________________________________

\section{Introduction}

Wolf-Rayet (WR) stars are the descendants of massive ($\ge$ 25 M$_\odot $) O-type stars and represent the last evolutionary phase of a massive stellar object prior to its explosion as supernova. WR stars are some of the most powerful sources of ionizing radiation and they have lost a significant portion of their atmospheres  through intense winds, which leads to the formation of the well known ``WR ring nebulae'' (WRRN)  \citep{chu81}    and/or ``interstellar bubbles'' (IB) \citep{c75,D77,W77}. WRRN and IBs are consistents with an evolution of an O-type star to the WR phase through a sequence of three stages. Along this evolutionary path, each stage is characterized by a different kind of wind \citep{gs95}.  During the O phase, the gas around the star is first  ionized by the high Lyman continuum flux, giving rise to an \hii region which expands in the surrounding cold neutral medium as a result of its higher pressure. Afterwards, the \hii region is evacuated via powerful stellar winds creating an IB. IBs were succesfully detected mainly in the 21 cm line of atomic hydrogen \citep{ar92,ac96,cnhk96,acrc99,vcm05}. When  the star becomes a red super giant (RSG), the stellar wind is dense and has a low terminal velocity. Afterwards, the WR phase begins, and their fast wind rapidly reaches and interacts with the previous RSG wind creating a WRRN. 
\begin{figure*}
\centering
\includegraphics[width=450pt]{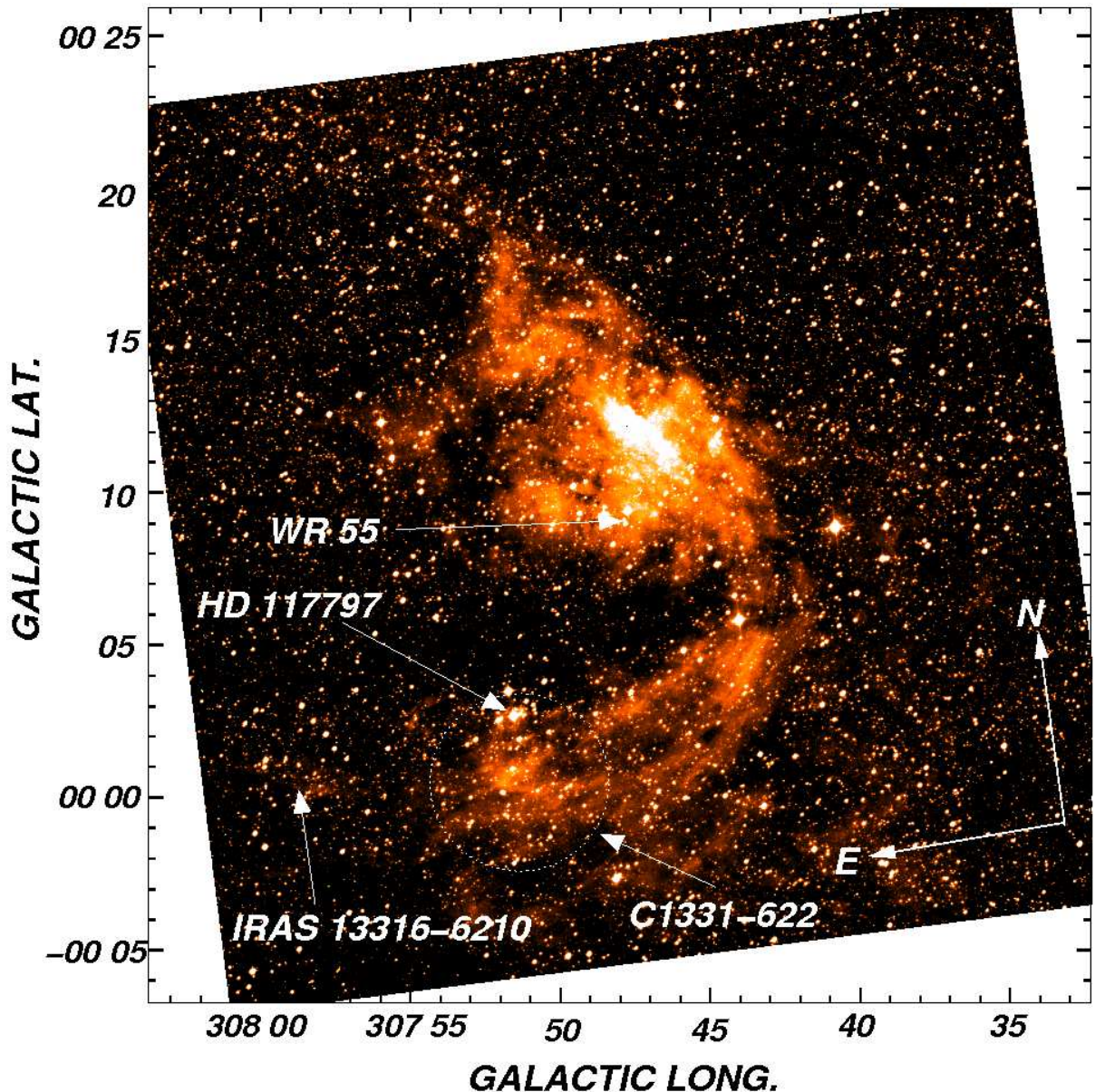}
\caption{SuperCOSMOS H${\alpha}$ image of the brightest part of the  ring nebula RCW 78. The position of different objects in the field are indicated (see text). The orientation of the equatorial system is given by the arrows labeled E (east) and N (north).   }
\label{fig:wr55halfa}
\end{figure*}

The physics and kinematics of molecular gas around ring nebulae and IBs are issues far from being  completely  understood. During the expansion of the \hii region, a dense shell of neutral material accumulates between the ionization front (IF) and  the shock front (SF), that in  an ideal case completely surrounds the \hii region or the IB. In most of  the cases studied so far, the molecular gas is interstellar in origin, and shows signs of interaction with the stellar radiation and winds   \citep{rmh01a,rmm01b,crg01,m01,vcp09}.

RCW 78  is a WRRN which was first noticed in the catalogue of H${\alpha}$ emission regions of the Southern Milky Way \citep{rcw60}. The brightest part of this nebula  is about 15$'$ in diameter  and is centered on the position of the star HD 117688 (Fig. {\ref{fig:wr55halfa}}). HD 117688 ($\equiv$ WR 55) is a WN7 star located at ({\it l,b}) = (307$^{\circ}$.80, +0$^{\circ}$.160)  \citep{vdh01} and the H${\alpha}$\ ring  nebula RCW 78 is likely to be associated with this star.    About 7$'$ southeast of WR 55 (in equatorial coordinates) the open cluster C1331-622 is located. This cluster is $\sim$ 7$'$ in diameter and is  820 pc \citep{d0210} away from the Sun. The O8Ib(f) star \citep{w82} HD 117797, located at ({\it l,b}) = (307$^{\circ}$.8593,+0$^{\circ}$.0447), is seen projected onto C1331-622. This star is located at a distance of 3.9  $\pm$ 1.0 kpc \citep{tf05}  and is unlikely to be related to the open cluster C1331-622. The error in the distance of HD 117797 stems  from assuming a cosmic dispersion of 0.5  in absolute magnitude \citep{w72}.    The location of HD 117797  prevents us from  determining whether the intense H$\alpha$ emission observed towards lower galactic longitudes  is caused by the presence of nebular emission associated with this star or is part of RCW 78.  The   IRAS source   \hbox{13316-6210} is situated almost $\sim$5$'$ eastwards from HD 117797.

 The distance of WR 55 and its associated nebula is far from being known. On the one hand, several authors quote distances in the range from 4.0 kpc to 5.0 kpc \citep{g88,crmr09}, while others authors have determined distances mostly in the range from 5.5 kpc to 7.6 kpc \citep{ct81,cv90,vdh01}

Several spectroscopic studies of RCW 78 were carried out in the past. \citet{ct81}, showed that the H${\alpha}$ line at the brightest central part of the nebula displays a slight north-south velocity gradient. The velocity of the H${\alpha}$ line varies from $-$44 \kms (at the position of WR 55)  to $-$53.4 \kms ($\sim$ 7$'$ northwards of WR 55). These authors claimed that this gradient is the result of an outflow of the ionized gas at the surface of a molecular cloud.  They classified RCW 78 as R$_a$-type nebula (amorphous) since no signs of expansion are found. Spatially resolved spectroscopy was carried out by \citet{evme90}. The derived abundances of nitrogen and helium are consistent with a  self-enrichment. Later on, \citet{e93} claimed that photoionization is the main source of excitation of the nebula. $^{12}$CO (J = 1$\rightarrow$0) and (J = 2$\rightarrow$1) lines using SEST data and  (J = 1$\rightarrow$0) line using NANTEN data were observed towards RCW 78  by \citet{crmr09}. They found a CO ring-like structure in the velocity range from  $-$52.5 to $-$43.5 \kms which is almost coincident with the brightest western part of the nebula. A  second CO structure in the velocity range from $-$43.5 to $-$39.5 \kms was also detected. The molecular gas in a region 12$'$ $\times$ 10$'$ centered on WR 55  shows similar characteristics in the line velocities to those of the ionized gas \citep{ct81}, i.e. larger negative velocities to the north, whilst more positive velocities close to the star. The authors suggest that WR 55 is not only responsible for the ionization  at the surface of the molecular cloud, but also for the shape and kinematics of RCW 78. Infrared MSX, IRAS, and GLIMPSE data were also analyzed by these authors.  They reported  two well separated spots centered at about  RA,Dec(J2000) = (13$^h$34$^m$15.0, $-$62$^\circ$26$'$) ({\it l,b} =  307$^{\circ}$52$'$, 00$^{\circ}$02$'$) and RA,Dec(J2000) = (13$^h$35$^m$10, $-$62$^\circ$26$'$) ({\it l,b} = 307$^{\circ}$51$'$, 00$^{\circ}$02$'$), where  a high number of candidate YSOs are found. The positions of these spots are coincident with two bright extended infrared sources observed at 60 $\mu$m (referred in that work as source B and source C, respectively). They  also claim  that the expansion caused by WR 55 could have triggered the  star formation disclosed by the presence  of Young Stellar Objects (YSO) candidates observed in the direction of the molecular gas in the velocity range of $-56.5$ to $-39.5$ \kms.

Given the information above, it is worth further exploring the   physical and kinematical characteristics of the molecular gas in RCW 78. Is this nebula simply a non-expanding  H{\sc ii} region? Is WR 55 interacting with the molecular and/or ionized gas of the nebula?\  How can the velocity gradient observed in the ionized and molecular gas be explained?

In order to perform a thorough study of the distribution of the molecular and ionized gas of RCW 78, as well as their physical and kinematical properties, optimal fully sampled and  high velocity resolution $^{12}$CO (J = 1$\rightarrow$0) observations covering a region of $\sim$ 25$'$ $\times$ 25$'$ around the star WR 55 were carried out.  The molecular observations were analyzed in conjuction with archival data of both radiocontinuum at 1.4 and 4.85 GHz, and \Ha\ observations to account for the ionized gas.  The databases used in this work are outlined in Sect. 2, the results are described in Sect. 3, and the discussion in Sect. 4. Conclusions are presented in Sect. 5. This is the first of a series of papers aimed at studying the characteristics of molecular gas around Wolf-Rayet ring nebulae.

%__________________________________________________________________

\begin{figure*}
\centering
\includegraphics[width=480pt]{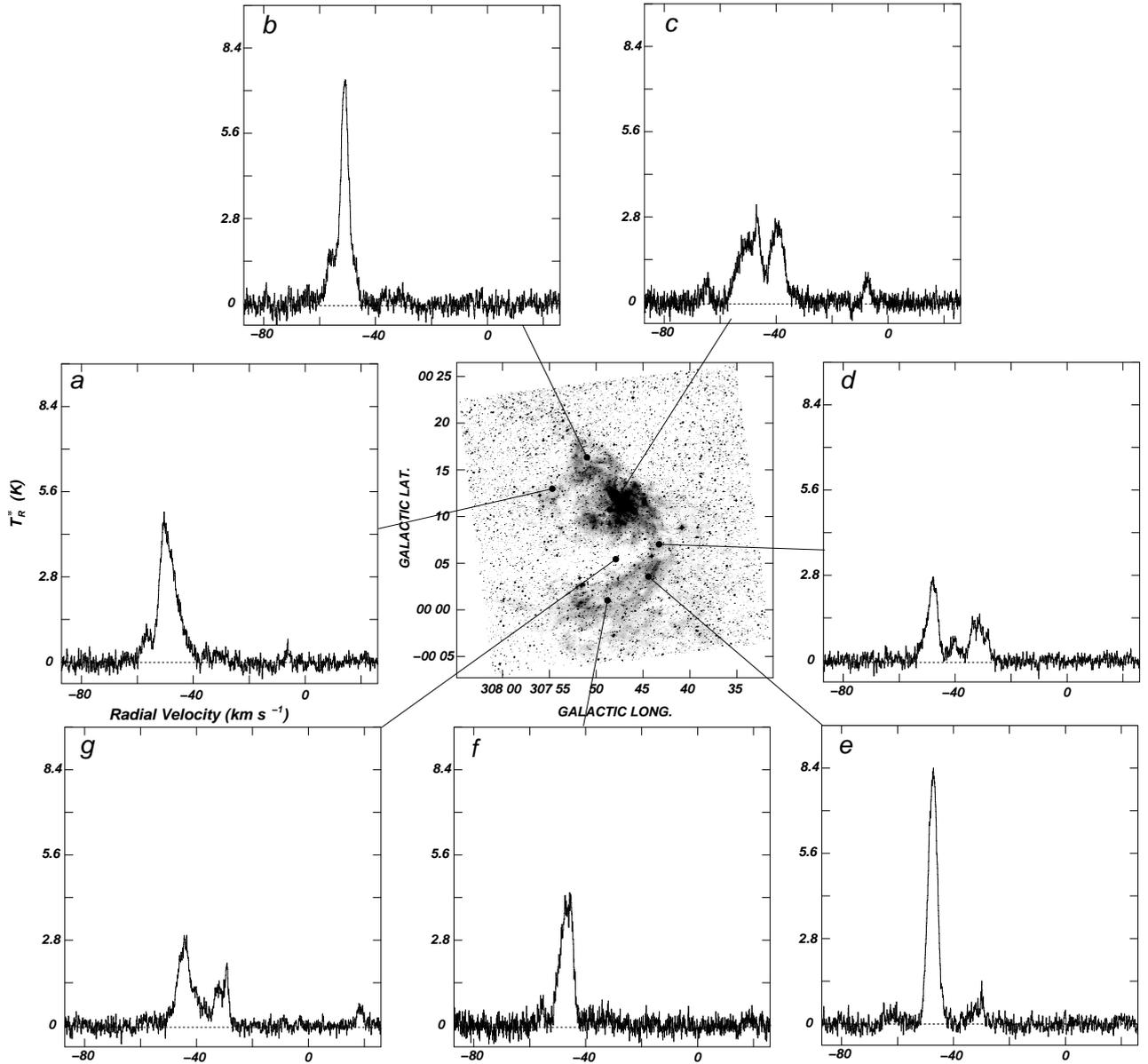}
\caption{ Mean CO emission profiles toward seven regions of the RCW 78 nebula around the star WR 55. The CO profiles are averaged over a square area $\sim$ 3$'$ in size, centered on the black dots drawn on the H${\alpha}$ image (center). The profile units are T$^{*}_R$ in K (ordinate) and V$_{LSR}$ in \kms.}
\label{fig:wr55ispec}
\end{figure*}

\section{Observations and data reduction}

The databases used in this work are:

\begin{enumerate}

\item Intermediate angular resolution, medium sensitivity, and
high-velocity resolution {\rm $^{12}$CO (J=1$\rightarrow$0)} data
 obtained with the 4-m {\rm NANTEN} millimeter-wave telescope of Nagoya University. At the time the authors carried out the observations, April 2001, this telescope was
 installed at Las Campanas Observatory, 
Chile. The half-power beamwidth and the system temperature, including the
atmospheric contribution towards the zenith, were  2$'$.6
 ($\sim$3.8 pc at 5 kpc) and $\sim$ 220\,K
 (SSB) at 115 GHz, respectively. The data were gathered using the position switching mode. Observations of points devoid of CO emission were interspersed among the
 program positions. The coordinates of these points were retrieved from a
database that was kindly made available to us by the NANTEN staff. The
 spectrometer used was an acusto-optical with 2048 
channels providing a velocity resolution of $\sim$ 0.055 \kms. For
intensity calibrations, a room-temperature chopper wheel was employed \citep{pb73}. 
An absolute  intensity calibration 
 \citep{uh76,ku81} was achieved by observing
Orion {\rm KL} (RA(1950.0)=5$^h$\,32$^m$\,47$^s$.0, 
Dec (1950.0) = $-$5$^\circ$\,24$'$\,21$''$), and $\rho$ Oph East  (RA(1950.0) =16$^h$ \,29$^m$ \,20$^s$9, Dec 
(1950.0) = $-$24$^\circ$\,22$'$,13$''$). The absolute radiation
 temperatures, T$_R^\ast$, of
Orion {\rm KL} and $\rho$ Oph East, as observed by the {\rm NANTEN} 
radiotelescope were assumed to be 65 K and 15 K, respectively \citep{myom01}. 
The {\rm CO} observations  covered a region ($\bigtriangleup${\it l}
 $\times$ $\bigtriangleup${\it b}) of 86$'$.4$\times$86$'$.4 centred at 
({\it l,b})=(307\gra.8, 0\gra.16) and the observed grid consists of points located  every 1\arcmin.35 (full sampling). A total of 489 positions were observed.Typically, the integration time per point was 16s resulting in an rms noise of $\sim$0.3 K. A second-order degree polynomial was substracted from the observations to account for instrumental baseline effects. The spectra were reduced using CLASS software (GILDAS working group)\footnote{\it htto://www.iram.fr/IRAMFR/PDB/class/class.html}.  \\

\item Narrow-band H${\alpha}$ data retrieved from SuperCOSMOS H-alpha Survey (SHS)  \footnote{{\it http://www-wfau.roe.ac.uk/sss/halpha/index.html}}. The images have a sensitivity of 5 Rayleigh,  %(1 Rayleigh = 2.41 $\times$ 10$^{-7}$ erg cm$^{-2}$ s$^{-1}$ sr$^{-1}$), 
 and $\sim$1$''$ spatial resolution \citep{p05}\\

\item Radio continuum observations:

\begin{itemize}

\item 1.4 GHz data retrieved from the Southern     Galactic Plane     Survey (SGPS)\footnote{{\it http://www.atnf.csiro.au/research/HI/sgps/queryForm.html}}.  The angular resolution is 100$''$ and the rms sensivity is below  1 \mjyb  \citep{hav06}.

\item 4.85 GHz data  retrieved from the Parkes-MIT-NRAO (PMN) Southern Radio Survey.   The images have $\sim$5$'$ resolution and $\sim$8 \mjyb rms noise plus confusion   \citep{c93}. 

\item 843 MHz image retrieved from the Molonglo Galactic Plane Survey (MGPS)\footnote{{\it http://www.astrop.physics.usyd.edu.au/MGPS/ }}. The angular resolution is \hbox{43$''$ x 43$''$cosec(dec)} and the rms sensitivity is 1-2 \mjyb.\\
\end{itemize}

\item Infrared data at 24 $\mu$m  (angular resolution $\sim$5$''$) obtained from the Multiband Imaging Photometer for {\it Spitzer} (MIPS) from the MIPS Inner  Galactic Plane Survey (MIPSGAL)\footnote{{\it http://sha.ipac.caltech.edu/applications/Spitzer/SHA//}} \citep{ca05}.

\end{enumerate}

\section{Results and analysis of the observations}

\subsection{CO emission}

 Representative CO profiles obtained at different positions towards RCW 78    are shown in Fig. {\ref{fig:wr55ispec}}.  The CO emission shows three main  velocity components: {\it i)} $\sim$ $-$58 to $-$46 \kms (profiles {\it a} to {\it g}), {\it ii)} $\sim$ $-$44 to $-$38 \kms (profile {\it c}, and probably {\it d}),  and {\it iii)} $\sim$ $-$33 to $-$28 \kms  (profiles {\it d}, {\it e}, and {\it g}). 

\begin{figure*}
\centering
\includegraphics[width=400pt]{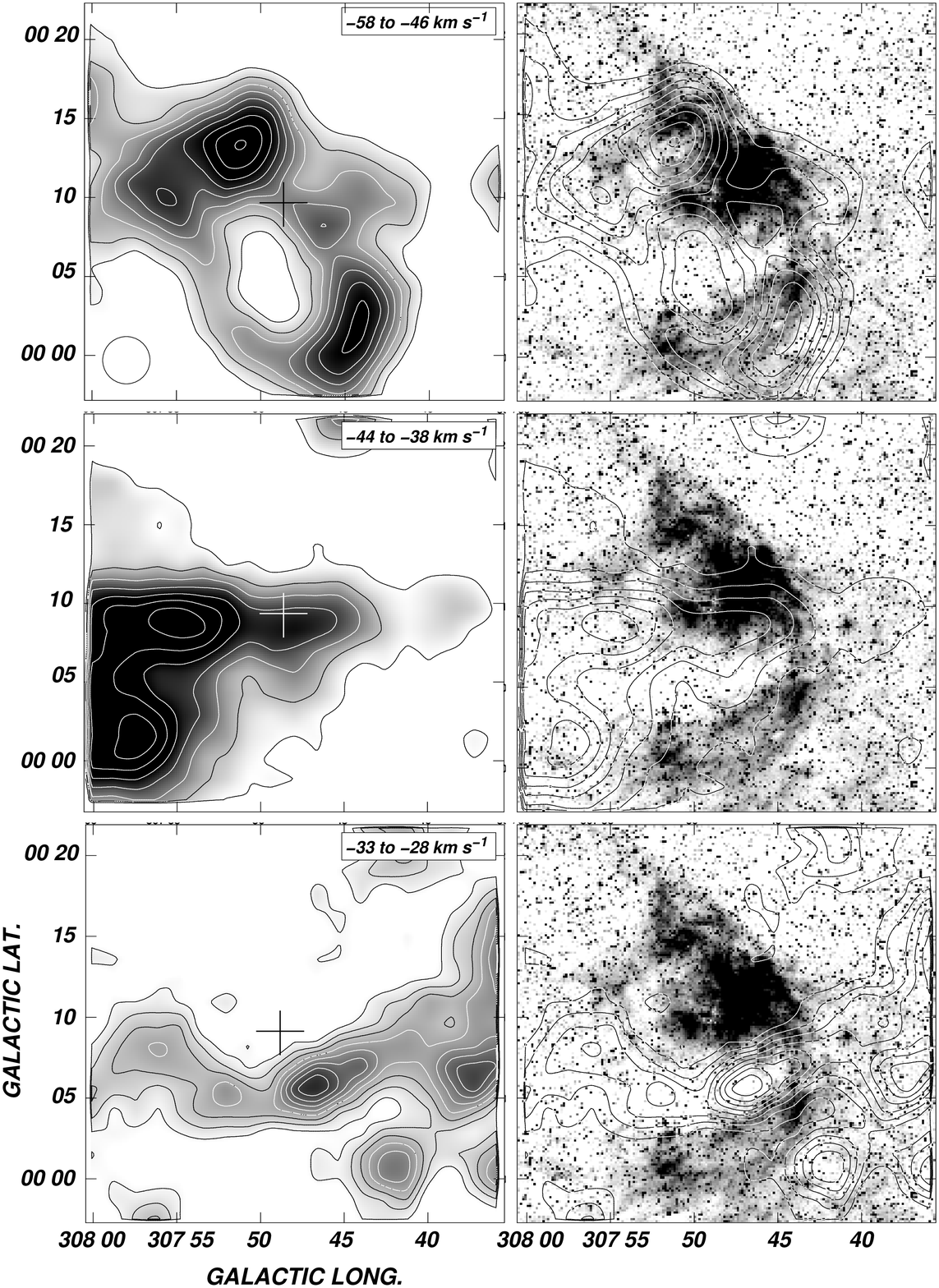}
\caption{ {\it Left panels}: Spatial distribution of CO emission in three velocity ranges. The velocity range is given in the upper right corner of each map. The line temperature is averaged over the corresponding velocity range. In the velocity range $\sim$ $-$58 to $-$46 \kms, the lowest contour is 0.5 K  ($\sim$ 20 rms) and the contour spacing is 0.35 K. In the velocity range $-$44 to $-$38 \kms, the lowest contour is 0.35 K  ($\sim$14 rms) and the contour spacing is 0.7 K. In the velocity range $-$33 to $-$28 \kms, the lowest contour is 0.42 K  ($\sim$ 11 rms) and the contour spacing is 0.21 K. In all cases the greyscale goes from 0.35 to 3.15 K. The position of WR 55 is marked by a plus sign at the center of each image. The beam size of the CO observations is shown by a  circle  in the upper left panel. {\it Right panels}:  Overlay of the mean CO emission (contours) in the three velocity ranges and the SHS H${\alpha}$ emission of RCW 78 (grey scale).    }
\label{fig:rcw783comp}
\end{figure*}

It is noticeable from Fig. \ref{fig:rcw783comp} that the molecular gas distribution is quite dissimilar among the three velocity ranges mentioned above. In the velocity range from $-$58 to $-$46 \kms  (upper left panel of Fig. \ref{fig:rcw783comp}) the molecular gas displays a very clumpy ring-like structure which has an excellent morphologic correlation with the H${\alpha}$ nebula, except for the lane of molecular gas betwen ({\it l,b}) $\approx$ (307$^{\circ}$54\arcmin, +00$^{\circ}$09\arcmin) and ({\it l,b}) $\approx$ (307$^{\circ}$52\arcmin, $-$00$^{\circ}$02\arcmin)  (see Fig. \ref{fig:wr55halfa} and Fig. \ref{fig:rcw783comp}, right panel) that has no \ha\ counterpart.  The velocities of the CO line   are similar to those observed at H${\alpha}$ by \citet{ct81}. This molecular structure was reported by \citet{crmr09}.  In the second velocity range,  $-$44 to $-$38 \kms (middle left panel of Fig. \ref{fig:rcw783comp}), the CO emission is mostly confined to a region whithin  307$^{\circ}$45\arcmin  $<$ $l$ $<$ 308$^{\circ}$00\arcmin  and  $-$00$^{\circ}$02\arcmin  $<$ $b$ $<$ +00$^{\circ}$12\arcmin. This molecular structure, also reported by \citet{crmr09}, shows  no resemblance with the \ha\  nebula. Along the third velocity range, $-$33 to $-$28 \kms (lower left panel of Fig. \ref{fig:rcw783comp}), the molecular gas appears   as an elongated feature mostly confined along  $b$ = +00$^{\circ}$05\arcmin

\begin{figure*}
\centering
\includegraphics[width=400pt]{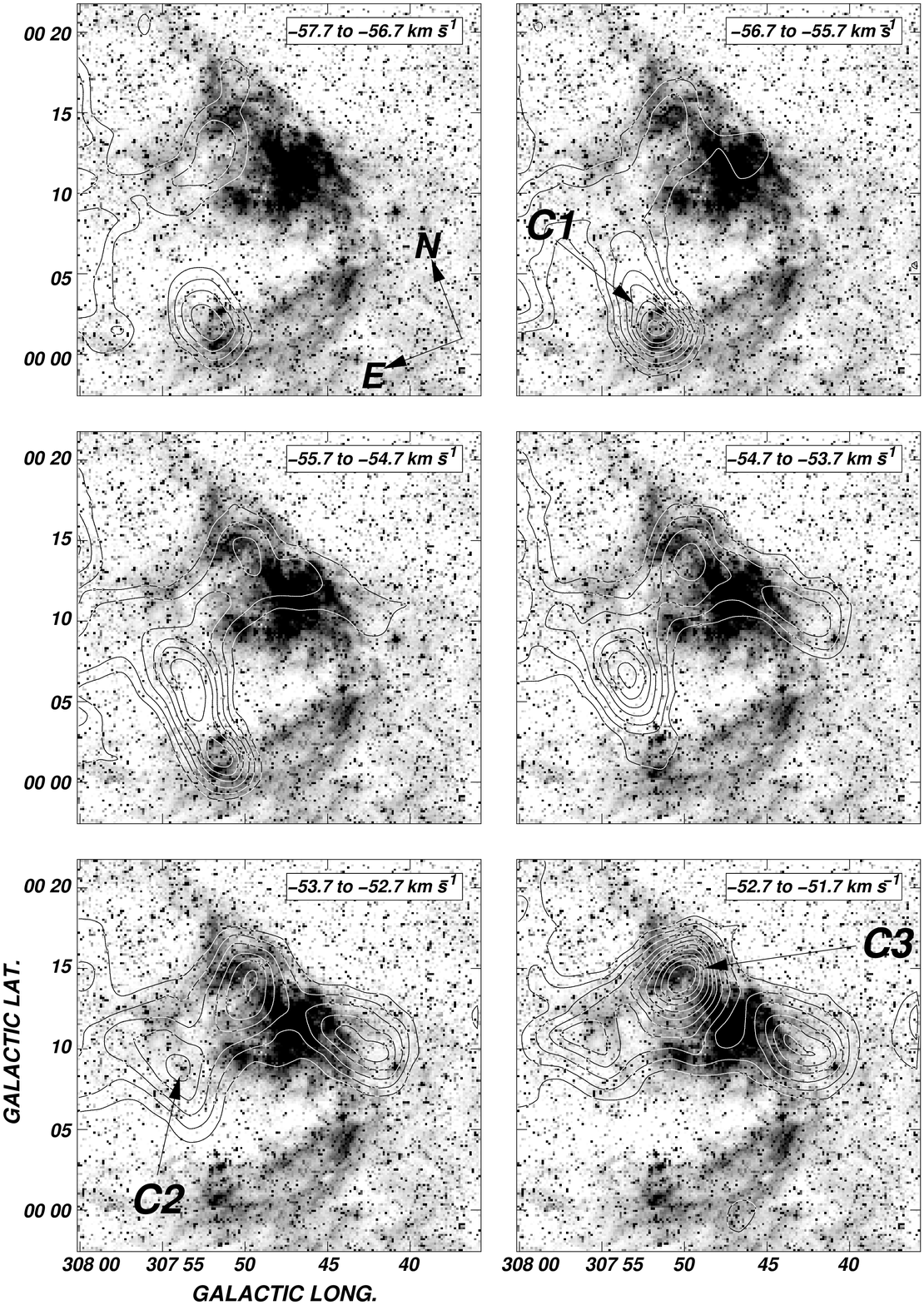}
\caption{Overlay of the mean CO emission (contours) in the velocity range from $\sim$ $-$57.7 to $-$45.7 \kms and the SHS H${\alpha}$ emission of RCW 78 (grey scale). The velocity interval of each map is indicated in the upper right corner of each image. The orientation of the equatorial coordinate system is given by the thick arrows labeled E (east) and N (north) at the first map. The lowest temperature contour is 0.8 K ($\sim$10 rms). The contour spacing temperature is 0.6 K.    }
\label{fig:rcw78mosaico}
\end{figure*}

\begin{figure*}
\centering
\includegraphics[width=400pt]{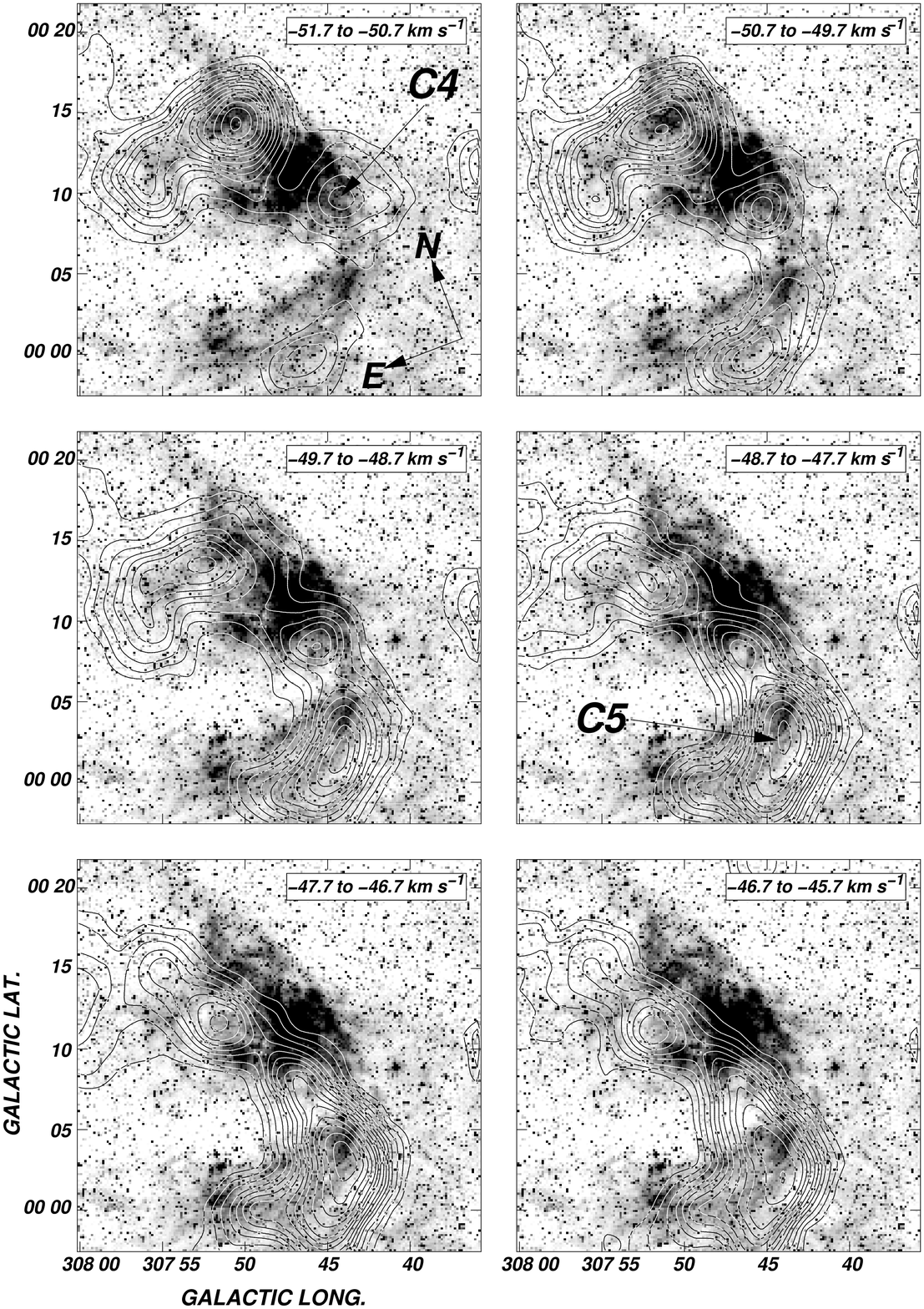}
\addtocounter{figure}{-1}
\caption{(continuation)}
\label{}
\end{figure*}

For the three molecular components, mean radial velocities weighted by line temperature ($\bar{V}$)   were derived using
\begin{eqnarray}
\qquad \qquad     \bar{V} \ =\  \frac{\sum_{i}\  T_{Peak_i}\ \times\  V_{Peak_i}}{\sum_{i}\  T_{Peak_i}}
 \label{eq:masaH2a}
\end{eqnarray}
where $T_{\rm Peak_i}$ and $V_{\rm Peak_i}$ are the peak temperature and the peak radial velocity of the $i$-spectrum obtained within the  first contour level used in Fig. \ref{fig:rcw783comp} for each molecular component.  The mean weighted radial velocities for the molecular gas in the three velocity ranges depicted in Fig. \ref{fig:rcw783comp} are   --49.5 \kms, --40.7  \kms, and -29.3 \kms, respectively. Using $\bar{V}$ and  the Galactic velocity field of \citet{bb93} along \hbox{{\it l} = 307\gra50\arcmin}, kinematical distances for the molecular features were derived. It ought to be pointed out that along this galactic longitude radial velocities more negative than  $-$47.5 \kms\ are forbidden. Nonetheless, it is  well established that noncircular motions of the order of \hbox{$\sim$ 8 \kms} \citep{bg78} are known to exist in the Galaxy.  Keeping this  in mind,  the molecular component at --49.5 \kms was assumed to be  in the neighborhood of the tangential point at a distance of $\sim$ 5 kpc. This distance is in agreement with one of the quoted distance ranges of WR 55. Based on this distance agreement, and   the excellent  correspondence in morphology and line velocities between the CO and H$\alpha$\ emission, we believe that this molecular component is very likely  related to both WR 55 and RCW 78.  For the molecular component at --40.7  \kms, near and far kinematical distances of  3.5 kpc and 6.9 kpc are derived. In Fig. \ref{fig:rcw783comp} (middle right panel) an overlay of this component with the \ha\ emission of RCW 78 is shown.  Though the  southern border of the CO feature appears to delineate very well the inner boundary of the \ha\, providing evidence in favour of a possible relationship, we suggest that this feature is very likely a background object with respect to RCW 78. The visual extinction (A$_v$) at ({\it l,b}) $\approx$ (307$^{\circ}$47\arcmin, +00$^{\circ}$09\arcmin) can be obtained using the equation of  \citet{b78} 
\begin{equation}\label{eq:dos}
  N(\rm H_2)/A_v = 0.94 \times 10^{21} \\ ({\rm cm^{-2} mag^{-1}})
\end{equation}
were $N({\rm H_2})$ is the molecular hydrogen column density (see below). We obtain in this direction a value of $N({\rm H_2})$ $\sim$ 4 $\times$ 10$^{21}$ molecules cm$^{-2}$, which yields to A$_v$ $\sim$ 4 magnitudes. Bearing in mind this figure, the visual absorption that would arise from the molecular gas do not diminish the brightness of the optical nebula in the way one would expect. Therefore, we suggest this object is unrelated to RCW 78 and it is located at the far kinematical distance. On the other hand, the CO feature at  --29.3 \hbox{\kms} (Fig. \ref{fig:rcw783comp}, lower panel)  shows an excellent spatial correlation with the high optical absorption lane seen at {\it b} $\approx$ +00$^{\circ}$05\arcmin\ between 307$^{\circ}$44$'$ $<$   $l$  $<$  307$^{\circ}$55$'$. Based on this, we suggest this feature is a foreground object with respect to RCW 78, locating it at its near kinematical distance ($\sim$ 2.8 kpc). Based on  the above, from here onwards we shall only concentrate on the analysis of the molecular gas distribution observed between --58 and --44 \kms, which is the only one likely to be  associated with RCW 78.

\begin{table*}
 \caption{Main physical parameters of the molecular concentrations C1, C2, C3, C4, and C5}
 \label{table:prop}      
 \begin{center}                        
 \begin{tabular}{l c c c c c}       
 \hline\hline                 
 %&&&&&\\             
Parameter     & C1 & C2 &  C3 & C4 & C5    \\
%    &&&&&\\
 \hline 
 
 Angular size ($'$)  &  $\sim$ 5  &  $\sim$ 10  &  $\sim$ 9   &   $\sim$ 7 &  $\sim$ 13  \\
 Linear size (pc)    &  5.6$^{\dag}$   &    14.5$^{\ddag}$   &  13$^{\ddag}$   &  10.2$^{\ddag}$  &  19$^{\ddag}$  \\
 $\bar{V}$ (\kms)   & $-$56.1 $\pm$ 0.2 & $-$52.8 $\pm$ 1.6 & $-$49.8 $\pm$ 1.3 & $-$49.4 $\pm$ 1.2   &  $-$47.1 $\pm$ 0.9  \\
%&&&&&\\
 $\bar{\Delta V}$ (\kms)  & 2.8 $\pm$ 0.5  &  5.1  $\pm$ 1.1   & 8.2 $\pm$ 2.3  &  5.1  $\pm$ 1.1  & 4.7 $\pm$ 0.9  \\
%&&&&&\\
 $\bar{T}_{\rm peak}$ (K)  &  3.1 $\pm$ 0.9   & 3.9 $\pm$ 0.5  & 4.8 $\pm$ 1.2  & 4.1 $\pm$ 0.8  & 5.4 $\pm$ 1.1  \\
% &&&&&\\

    $N_{\rm H_2}$  (10$^{21}$ cm$^{-2}$)& 0.9 $\pm$ 0.2 & 3.3 $\pm$ 0.5    &  4.0  $\pm$ 0.6 & 2.2 $\pm$ 0.3  &   3.9 $\pm$ 0.7      \\
%   &&&&&\\
$M_{\rm tot}$ (10$^3$ M$_{\odot}$)&  0.6 $\pm$ 0.2$^{\dag}$  & 6.8 $\pm$ 2.9$^{\ddag}$  &  8.7 $\pm$ 3.4$^{\ddag}$  &  4.2 $\pm$ 1.6$^{\ddag}$ & 13.8 $\pm$ 5.5$^{\ddag}$      \\ 

$n_{\rm H_2}$ (cm$^{-3}$) & 16 $\pm$ 5  & 118 $\pm$ 50   &  108  $\pm$ 42  &  91 $\pm$ 34  & 101 $\pm$ 41     \\

$\bar{T}_{\rm exc}$ (K)  &  7.6   &  8.9    &  10.2    & 9.4  & 11.1    \\

\hline                                   
\end{tabular}
\end{center}

\hspace{0.1cm} {{\scriptsize $^{\dag}$ considering a distance of 3.9 kpc (see Sect 4.1)}}

\hspace{0.1cm}   {{\scriptsize $^{\ddag}$ considering a distance of 5 kpc (see Sect. 3.1}}
\end{table*}

In Fig. \ref{fig:rcw78mosaico} a collection of narrow velocity images depicting the CO spatial distribution in the velocity  range from $-$57.7 to $-$45.7  \kms is shown. Every image represents an average of the CO emission over a velocity interval of 1 \kms (20 individual channel maps). The CO emission distribution shown in Fig. \ref{fig:rcw78mosaico} (in contours) is  projected onto the SHS H${\alpha}$ image of RCW 78 (greyscale). The velocity interval of the individual images is indicated in the upper   right corner.

For the sake of further analysis, five molecular concentrations are identified. They are labelled from C1 to C5 in order of increasing radial velocity. Concentration C1, whose maximum is at \hbox{({\it l,b}) $\approx$ (307$^{\circ}$53\arcmin, 00$^{\circ}$03\arcmin)},  is visible from $-$57.7 to $-$54.7  \kms attaining a maximum emission temperature of $\sim$ 4.8 K. This feature appears projected onto the region where HD 117797 and the open cluster C1331-622 are located. This region is also coincident with the area where candidate YSOs were reported by \citet{crmr09}.   At slightly more positive velocities, concentration C2 is found. This feature, is first noticed in the velocity interval from  $-$56.7 to $-$55.7 \kms, as an  extension of C1, along a position angle of $\sim$  45$^{\circ}$. Towards more positive velocities, the peak emission of this feature is shifted towards slightly higher longitudes and latitudes. In the velocity range from $-$52.7 to $-$48.7 \kms, C2 has a very good spatial correlation with the faint H${\alpha}$ emission seen in the northeastern  part of RCW 78, at $l$ $>$ 307$^{\circ}$50\arcmin. Concentration C3, spans the velocity range from  $-$55.7 to $-$48.7 \kms and has a very good morphological correlation with the H${\alpha}$ emission of RCW 78 seen 5$'$ northwards of WR 55. This component reaches a maximum emission temperature of 7.7 K at \hbox{({\it l,b}) $\approx$ (307$^{\circ}$50\arcmin, 00$^{\circ}$14\arcmin)} in the velocity range from  $-$51.7 to $-$50.7 \kms.  Concentration C4 becomes first noticeable in the velocity range from $-$55.7 to $-$54.7 km $^{-1}$ as a weak  feature seen projected onto the low galactic longitude extreme of RCW 78. This feature remains clearly visible as a separate feature till $-$47.7 km $^{-1}$. At more positive velocities this concentration is very difficult to follow because it merges with the northernmost extreme of C5. The latter  is first observed in the velocity range   $-$51.7 to $-$50.7 km $^{-1}$ as a detached CO emission feature seen projected slightly offset from  the \ha\  southernmost extreme of RCW 78. Concentration C5 reaches a maximum temperature of 8.9 K at at \hbox{({\it l,b}) $\approx$ (307$^{\circ}$44\arcmin, 00$^{\circ}$03\arcmin)}  in the velocity interval from  $-$47.7 to $-$46.5 \kms.
\begin{figure}[h!]
\centering
\includegraphics[width=253pt]{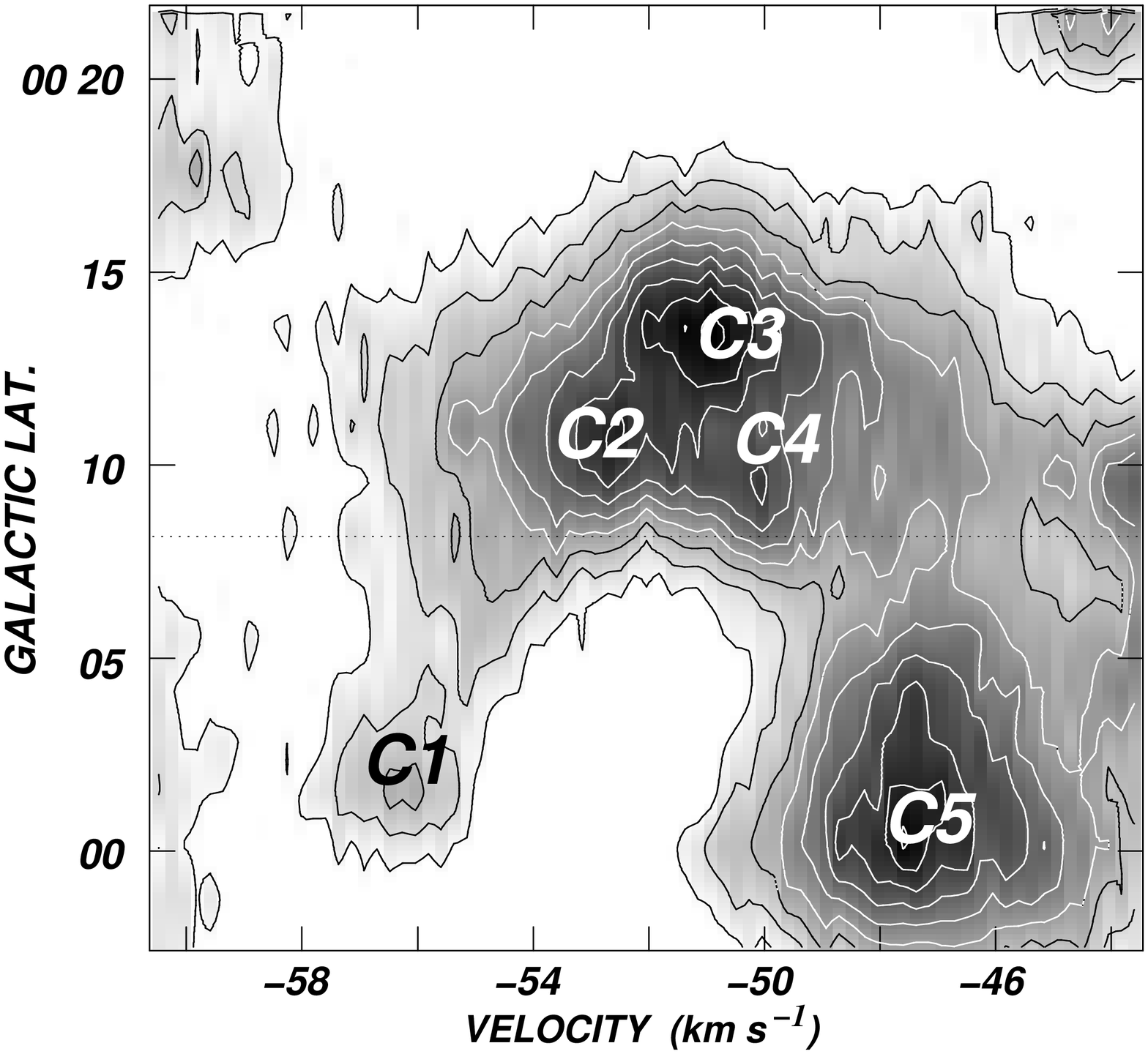}
\caption{ Galactic latitude-velocity contour map of the average CO emission in the galactic longitud range  from 307\fdg67 to 308\fdg00. The lowest temperature contour is 0.5 ($\sim$3.5 rms). The contour sapcing is $\sim$0.3 K. The location of WR 55 is indicated by the dotted line.  }
\label{fig:rcw78-bv}
\end{figure}

 In Table \ref{table:prop} we list some physical and geometrical properties of the molecular concentrations C1 to C5. The angular and linear sizes of the concentrations are listed in rows 1 and 2, respectively. Figure \ref{fig:rcw78mosaico}    shows that a velocity gradient is present  among the five molecular concentrations. In Table \ref{table:prop}   the averaged velocity ($\bar{V}$), width ($\bar{\Delta V}$), and peak temperature $\bar{T}_{peak}$, obtained from a gaussian fitting  to the mean CO profile of each molecular concentration, are given in rows 3, 4, and 5, respectively. To obtain these profiles all individual CO spectra within the 0.5 K contour level in Fig. \ref{fig:rcw783comp} were added up and averaged.  Clearly, the mean velocity of the concentrations increases from C1 to C5.   This velocity gradient is also noticeable in the velocity-galactic longitude image shown in Fig. \ref{fig:rcw78-bv}.  This image shows the averaged  CO emission in the galactic longitud range  307$^{\circ}$42\arcmin $<$ $l$ $<$ 308$^{\circ}$00\arcmin.  Concentrations  C1, C2, C3, C4, and C5 are identified. 

The  mass of the molecular gas  was derived  making use of  the empirical relationship between the molecular hydrogen column density, $N(\rm H_2)$, and the integrated molecular emission, \hbox{$I_{{\rm ^{12}CO}}$ ($\equiv\int\ T^*_{R}   \ d{\rm v}$)}. The conversion between $I_{{\rm ^{12}CO}}$  and $N(\rm H_2$) is given by the equation 
\begin{equation}\label{eq:cero}
 \quad   N({\rm H_2})\ =\ (1.9\ \pm\ 0.3)\ \times\ 10^{20}\ I_{{ ^{12}CO}} \ \ \ \quad  ({\rm cm}^{-2})
\end{equation}
  \citep{ d96,sm96}. The total  molecular mass $M_{\rm tot}$, was calculated through
\begin{equation}\label{eq:uno}
  \quad  M_{\rm tot}\ =\  (m_{sun})^{-1}\  \mu\ m_H\ \sum\ \Omega\ N({\rm H_2})\ d^2 \qquad  \quad  \ \quad ({\rm M}_{\odot})
\end{equation}
where  $m_{sun}$ is the solar mass ($\sim$ 2 $\times$ 10$^{33}$ g),    $\mu$ is the mean molecular weight, assumed to be equal to 2.8 after allowance of a relative helium abundance of 25\% by mass \citep{Y99},  $m_{H}$ is the 
hydrogen atom mass   ($\sim$ 1.67 $\times$ 10$^{-24}$ g), $\Omega$ is the solid angle subtended by the CO feature  in ster, $d$ is the  distance, expressed in cm, and  $M_{\rm tot}$ is given in units of solar masses. Values of  $N({\rm H_2})$, $M_{\rm tot}$, and molecular volume density ($n_{H_2}$)  for each molecular concentration are quoted in rows 4, 5, and 6 of \hbox{Table \ref{table:prop}}, respectively.  
 
To probe the surface conditions of the molecular gas, mean excitation temperatures ($\bar{T}_{exc}$) were calculated.  Assuming $^{12}$CO emission to be optically thick,  averaged  excitation temperatures were obtained from 
\begin{equation}\label{eq:tpeak}
  \bar{T}_{\rm peak}\ (^{12}CO)\ =\ J_{\nu}(\bar{T}_{\rm exc})\ -\ J_{\nu}(T_{\rm bg})
\end{equation} 
\citep{d78} where $J_{\nu}$ is the Planck function at a frequency $\nu$, and $T_{bg}$ is the background temperature \hbox{($\sim$ 2.7 K)}. These values are  presented in row 7 of Table \hbox{\ref{table:prop}}.  It is worth noting that Equation \ref{eq:tpeak}    assumes a filling factor unity, which implies that the values of $T_{\rm exc}$ quoted in Table \hbox{\ref{table:prop} must be considered as lower limits.}

\begin{figure*}
\centering
\includegraphics[width=492pt]{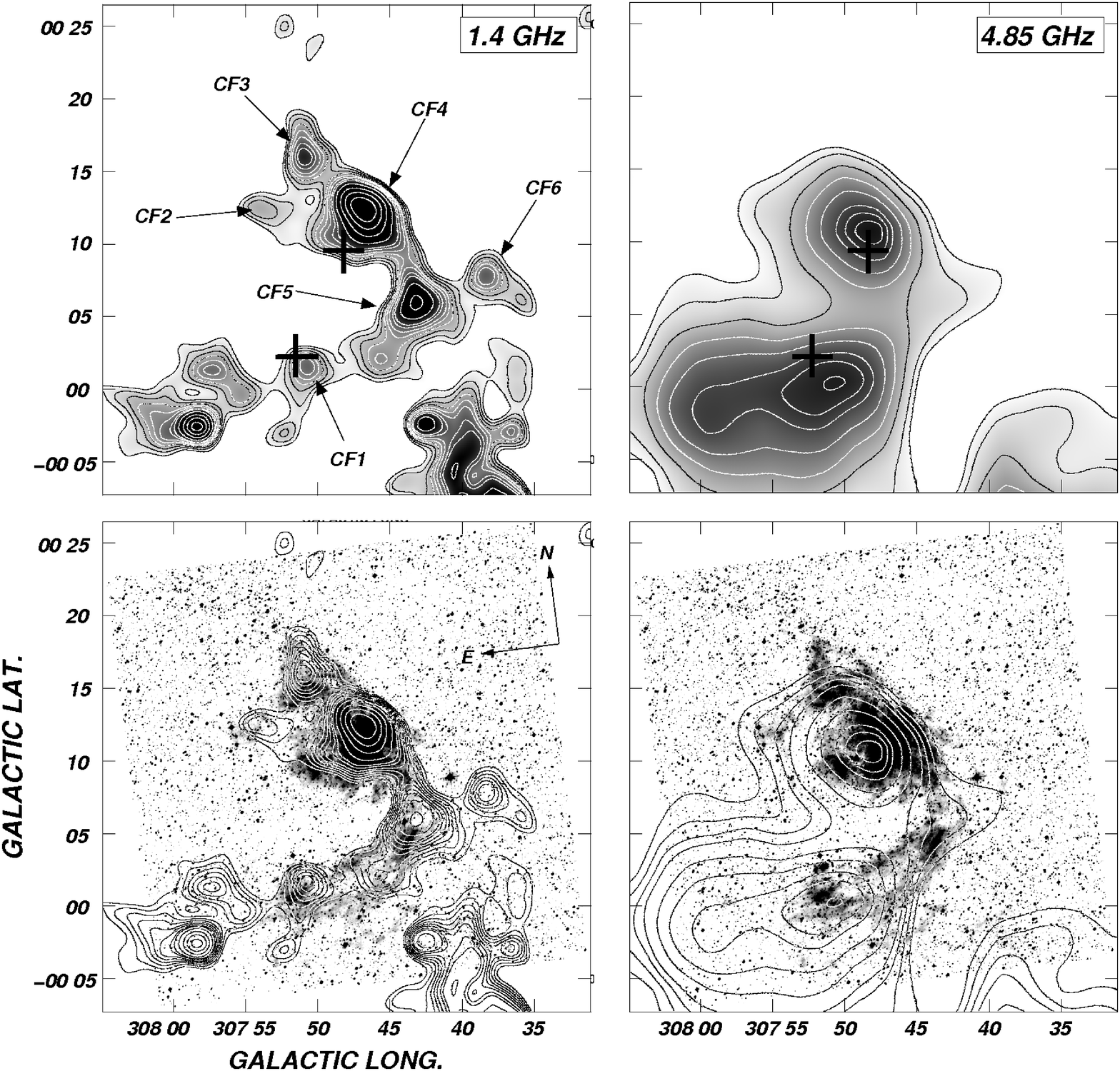}
\caption {{\it Upper panels:} Radio continuum emission distribution at 1.4 GHz (left panel) and 4.85 GHz (right panel). Contour levels for the 1.4 GHz image go from    21 mJy beam$^{-1}$  to  42 mJy beam$^{-1}$ in steps  of   7 mJy beam$^{-1}$ and from 60  mJy beam$^{-1}$ in steps of 10 mJy beam$^{-1}$. For the 4.85 GHz the lowest contour level is 40 mJy beam$^{-1}$ and the contour spacing is 20 mJy beam$^{-1}$. Black crosses indicate the positions of WR 55 and HD 117797. {\it Lower panel:} Overlay of the radiocontinuum emission at 1,4 GHz and 4.85 GHz  on the H$\alpha$ emission of RCW 78.    }
\label{fig:cont}
\end{figure*}

\subsection{Radio continuum emission}

The radio continuum emission distribution at 1.4 GHz obtained from the ATCA archives  is shown in the upper left panel of Fig. \ref{fig:cont}. Regardless of the difference in the angular resolution with the SuperCOSMOS image, radio continuum emission distribution resembles the H$\alpha$ emission of the nebula (see Fig. \ref{fig:cont}, left lower panel). Two intense features, having no counterpart with the optical emission of RCW 78, are seen toward the south and southeast of the optical  nebula. The southern feature, located in a region between 307\gra35\arcmin $<$ $l$ $<$ 307\gra45\arcmin\   and\ $-$0\gra06\arcmin $<$ $b$ $<$      0\gra02\arcmin\  is probably related to an intense radio continuum source located at  ({\it l,b}) = (307\gra37\arcmin, $-$0\gra18\arcmin) (not shown here), while the southeastern feature, located in a region between\    307\gra55\arcmin   $<$ $l$ $<$      308\gra05\arcmin\   and\ $-$0\gra05\arcmin\  $<$ $b$ $<$   0\gra02\arcmin\ is likely related to  the source IRAS13316-6210. These features will not be considered any further in our analysis.

Six  intense features peaking at about ({\it l,b}) = (307\gra51\arcmin, 0\gra02\arcmin), ({\it l,b}) = (307\gra54\arcmin, 0\gra12\arcmin),  ({\it l,b}) = (307\gra51\arcmin, 0\gra16\arcmin), ({\it l,b}) = (307\gra47\arcmin, 0\gra13\arcmin),  ({\it l,b}) = (307\gra44\arcmin, 0\gra06\arcmin), and ({\it l,b}) = (307\gra38\arcmin, 0\gra08\arcmin) are seen projected onto the optical nebula. Considering  their positions, in accordance with the molecular concentrations described in Sect. 3.1, these features will be referred to as  CF1, CF2, CF3, CF4, CF5 and CF6, respectively. In order to determine their radio continuum flux densities at 1.4 GHz (S$_{1.4}$), the emission distribution  was analysed making use of the NOD2 package \citep{h74}. The values of S$_{1.4}$ for each feature  were calculated, after substracting a  background emission, and were summarized in \hbox{Table \ref{tabla-4,85ghz}}. The error quoted for S$_{1.4}$  stems from the uncertainty of the   first contour level used to define the sources. 

\begin{table*}
\caption{ Radio continuum parameters of continuum features CF1, CF2, CF3, CF4,  CF5, and CF6 derived from the 1.4 GHz emission }
\label{tabla-4,85ghz}
\begin{center}
\begin{tabular}{ l  c  c  c c c c}
\hline\hline 
%& &  & & & &\\              
          &  CF1  &  CF2 &  CF3 & CF4 & CF5 & CF6 \\
%& &  & & & &\\
\hline
%& &  & & & &\\
 S$_{1.4}$   (mJy)           & 23 $\pm$ 7     &  21 $\pm$ 7   & 55 $\pm$ 8  & 287  $\pm$ 21      &  131 $\pm$ 25 & 27 $\pm$ 8     \\
%&   &  & & & &\\
Peak position ({\it l,b})    &  307\gra51\arcmin, 0\gra02\arcmin           &  307\gra54\arcmin, 0\gra12\arcmin       & 307\gra51\arcmin, 0\gra16\arcmin  & 307\gra47\arcmin, 0\gra13\arcmin &  307\gra44\arcmin, 0\gra06\arcmin & 307\gra38\arcmin, 0\gra08\arcmin   \\
%& & &\\
$T_{b}$ (K) &2.4   &  1.9 & 3.0   & 5.9  & 4.1 & 2.5\\
%&   &   &  & & &\\
$EM$  (10$^3$ pc cm$^6$)   &    1.5     &  1.2     &   1.9 & 3.7 & 2.6 &  1.6   \\
%& & & & & &\\
n$_e$ (cm$^3$)   & 21 $\pm$ 7  $^{\dag}$     & 20 $\pm$ 6$^{\ddag}$   &  21 $\pm$ 6$^{\ddag}$    & 20 $\pm$ 5$^{\ddag}$   & 22 $\pm$ 6$^{\ddag}$   & 20 $\pm$ 5$^{\ddag}$   \\

\hline

\end{tabular}
\end{center}

\hspace{0.1cm} {{\scriptsize $^{\ddag}$ considering a distance of 5 kpc (see text)}}

\hspace{0.1cm}   {{\scriptsize $^{\dag}$ considering a distance of 3.9 kpc (see Sect 4.1)}}

\end{table*}

To better  characterize the nature of  the radio continuum emission arising from RCW 78, the 4.85 GHz image obtained from the PMN Southern Radio Survey (Fig. \ref{fig:cont}, upper right panel) was also analysed. Though a detailed comparison between the 4.85 GHz and the  \Ha\ emission is  difficult due to the difference in angular resolution, two different features at 4.85 GHz, peaking at about ({\it l,b}) = (307\gra48\arcmin, 0\gra11\arcmin) and ({\it l,b}) = (307\gra51\arcmin, 0\gra01\arcmin),     seem to be morphologically correlated with the optical nebula. The first of them appears projected onto the brightest region of the optical nebula and its first contour level engulfs CF2, CF3, CF4, CF5, and CF6.  As expected, the intense emission at 4.85 GHz  arises from the central part of the nebula, toward the region of CF3 and CF4. The 4.85 GHz emission counterpart  of CF2 is the low brightness elongation seen  around ({\it l,b}) $\approx$ (307$^{\circ}$54\arcmin, +00$^{\circ}$13\arcmin), while the corresponding to CF5 and CF6  is the low intensity  sharp-pointed structure  observed around ({\it l,b}) $\approx$ (307$^{\circ}$43\arcmin, +00$^{\circ}$07\arcmin). The morphological correspondences described above   are better visualized after convolving the 1.4 GHz image down to the angular resolution of the 4.85 GHz image (not shown here).       The contiuum flux density  measured for this feature is  S$_{4.85}$ = 440 $\pm$ 50 mJy. To obtain the spectral index towards this region, the flux density at 1.4 GHz of features CF2, CF3, CF4, CF5, and CF6 were added up. These values provide an spectral index of $\alpha$ = $-$0.14 $\pm$ 0.05, which is compatible with the optically thin regime of an H{\sc ii} region.

The second feature detected at 4.85 GHz is seen projected onto HD 117797 and very likely represents the 4.85 GHz counterpart of CF1. Unfortunately, the angular resolution of these observations makes it almost impossible to isolate this feature from the near by emission structure peaking approximately at ({\it l,b}) $\approx$ (307$^{\circ}$58\arcmin, --00$^{\circ}$02\arcmin). This fact prevents us from deriving a reliable continuum flux density at this frequency. The radio continuum image at 843 MHz retrieved from MGPS (not shown here) do not show instrumental artifacts in the area of CF1. Using this survey we derived for CF1 at this frequency a flux density of 24 $\pm$ 5 mJy. Using this value and the 1.4 GHz flux density quoted from CF1 in Table {\ref{tabla-4,85ghz}}, a spectral index of \hbox{$\alpha$ = $-$0.08 $\pm$ 0.15} is derived.  Again, the spectral index is compatible,  within the error, with  free-free emission of an H{\sc ii} region in the optically thin regime. This speaks in favour of interpreting CF1 as the H{\sc ii} region created by the early type star HD 117797.

The emission measure ($EM$$= \int$ n$_e^2$  $dl$) of an H{\sc ii} region can be obtained via the relationship between optical depth at a frequency $\nu$ ($\tau_{\nu}$) and $T_b$ given by
\begin{eqnarray}
\quad   T_b =  T_e \times  (1-e^{-\tau_{\nu}}) \qquad \qquad \qquad \qquad \qquad (K)
\label{tb-tau}
\end{eqnarray}
were  $T_e$ is the electron temperature (considered to be $\sim$10$^4$ K), and the optical depth  ($\tau_{\nu}$)  is given by
\begin{eqnarray}
\quad   \tau_{\nu}\ =\ 0.08235\ T_e^{-1.35}\ \nu^{-2.1}\ EM 
\label{tau}
\end{eqnarray}
In Eq. {\ref{tau}}, $\nu$ is given in GHz and $EM$ in pc cm$^{-6}$. Using the radio continuum emission at 1.4 GHz, values of $EM$ were calculated for each feature. These values are shown in row 4  of Table {\ref{tabla-4,85ghz}}. Considering pure hydrogen plasma and adopting an extent along the line of sight equal to the observed minor axis, and using the $EM$ values determined before electron densities (n$_e$)  were calculated. These values are quoted in row 5 of Table {\ref{tabla-4,85ghz}}.

\begin{figure}[h!]
\centering
\includegraphics[width=250pt]{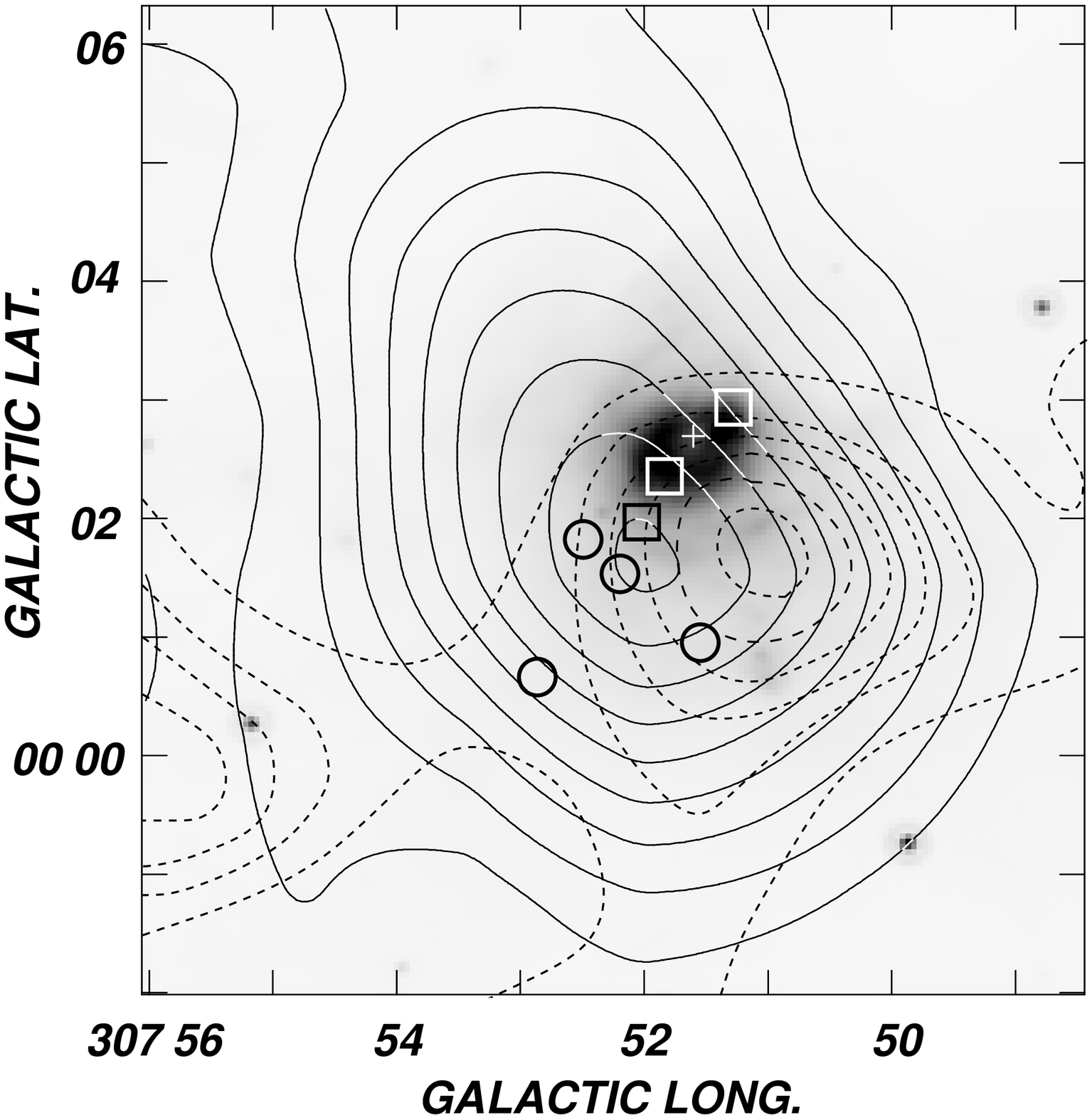}
\caption{ Overlay of the  MIPSGAL 24 $\mu$m emission (greyscale) superimposed on the mean CO emission in the velocity interval from $-$57.7 to $-$54.7 \kms (solid  contours), and the 1.4 GHz emission (dashed contours).   The position of HD 117797 is indicated with the white cross. The MSX and GLIMPSE candidate YSOs reported by \citet{crmr09} are indicated with squares and circles, respectively.}
\label{fig:c1-ysos}
\end{figure}

\section{Discussion}

\subsection{Concentration C1}

A look at Table {\ref{table:prop}} shows that the mean CO line width ($\bar{\Delta V}$), the molecular column density ($N_{\rm H_2}$), the total mass   ($M_{\rm tot}$), and volume density ($n_{H_2}$)   of concentration C1 are systematically  lower than the corresponding physical parameters of the other CO concentrations. The fact that its mean line width is on the average almost half of the line width of the other concentrations may be indicative that its dynamical state is different. In Fig. \ref{fig:c1-ysos} the MIPSGAL 24 $\mu$m emission of a  $\sim$ 8$'$ $\times$ 8$'$ region  centered on HD 117797   superimposed on the mean CO emission in the velocity interval from $-$57.7 to $-$54.7 \kms (concentration C1) and radiocontinuum emission at 1.4 GHz of CF1  are shown. This image shows that the star HD 117797 and the \hii\  region CF1 are seen  projected onto the molecular concentration C1 and that HD 117797 appears projected close to the centre of CF1.  In the same figure a series of candidates YSOs reported by \citet{crmr09} are also seen  in projection onto C1. Based on above, we may speculate that C1 is a molecular concentration that: a) is being partially ionized by HD 117797, and b) may be experiencing a star-forming process, revealed by the presence of candidate YSOs.

In order to check the first assumption, the total number of Lyman continuum photons ($N_{\nu}$)  needed to keep  CF1 ionized will be calculated. This figure is given by
\begin{eqnarray}
N_{\nu} = 0.76\ \times\ 10^{47}\ T_{4}^{-0.45}\ \nu^{0.1}\ S_{\nu}\ d^2 
\label{lyman}
\end{eqnarray}
\citep{ch76}, where $T_{4}$ is the electron temperature in units of 10$^4$ K, $\nu$ is the frequency in units of GHz,   $d$ is the distance in kpc, and  $S_{\nu}$ is the total flux density in Jy.  Substituting in Eq. {\ref{lyman}} the appropiate values, $N_{\nu}$ $\approx$ 3 $\times$ 10$^{46}$ s$^{-1}$ is obtained. This number is a lower limit to the total number of Lyman continuum photons required to maintain the gas ionized, since about 50 $\%$ of the UV photons is absorbed by dust mixed with the gas  in the H{\sc ii} region \citep{i01}.  Therefore, the number of Lyman continuum photons needed to power CF1 could be provided by HD 117797, since this number is a small fraction of the total ionizig photons emitted by a O8I star \hbox{($\sim$ 1 $\times$ 10$^{48}$ s$^{-1}$)} \citep{ma05}. 
 
A  noticeable feature in Fig. \ref{fig:c1-ysos} is a ringlike structure  $\sim$ 1$'$ diameter centered at the position of HD 117797,  seen at 24 $\mu$m emission. The MIPSGAL image allowed us to conclude that the observed IR emission detected by \citet{crmr09} in this direction   (source ``B'')  arises from  this feature.

If HD 117797 is the powering source of the \hii region CF1 and the latter is associated with the molecular concentration C1, then the distance of this complex is the distance to the star, namely 3.9 $\pm$ 1.0 kpc. If the star WR 55 and its associated nebula RCW 78 were close to the values determined by \citet{ct81} and \citet{vdh01}, around 6 - 7.6 kpc, then there will be no relationship between HD 117797 and WR 55. In turn this implies that the molecular concentration C1 and the \hii region CF1 are not related to RCW 78 and its associated molecular and ionized gas. On the other hand, if the distance to WR 55 were close to the estimates of \citet{g88}, namely 4 kpc, a physical association with HD 117797 (and its associated molecular and ionized gas) could not be ruled out. More accurate distance determinations to WR 55 and HD 117797 are necessary to shed some light on this  issue.

\subsection{A simple scheme for the molecular gas associated with RCW 78}

In an attempt to explain the morphology and radial velocity of the remaining molecular concentrations (C2, C3, C4, and C5), a simple geometrical model was elaborated taking into account the following constraints: 1) the morphology of the molecular gas, 2) the velocity gradient exhibited by the molecular gas, 3) the angular distribution of the different CO concentrations, and 4) the morphology of the H$\alpha$ emission. 

Because massive stars are born deeply buried within dense molecular clouds,  the classical IB scenario predicts that the molecular gas should expand spherically around the star. %If molecular shells  are two-dimensional projections of expanding spherical shells,   a connected structure with the velocity would be manifested in a data cube, first as a blueshifted pole, then as a growing and decreasing circular  ring, and finally as a redshifted pole, with the powering star placed at the center or close to it. 
Under the assumption  of a spherically symmetric expansion, a shell having a central velocity V$_0$ and an expansion velocity V$_{exp}$ should depict in a position-position diagram a ``disk-ring'' pattern when observed at different velocities. At V$_0$ the shell should attain its maximum diameter while at extreme velocities (either approaching or receding) the molecular emission should look like a disk. At intermediate velocities the radious of the ring shrinks as the extreme velocity are approaching.  According to Fig. \ref{fig:rcw78mosaico}, this behaviour is not observed in our CO observations, which disagrees with  the IB model.

\begin{figure}[h!]
\centering
\includegraphics[width=260pt]{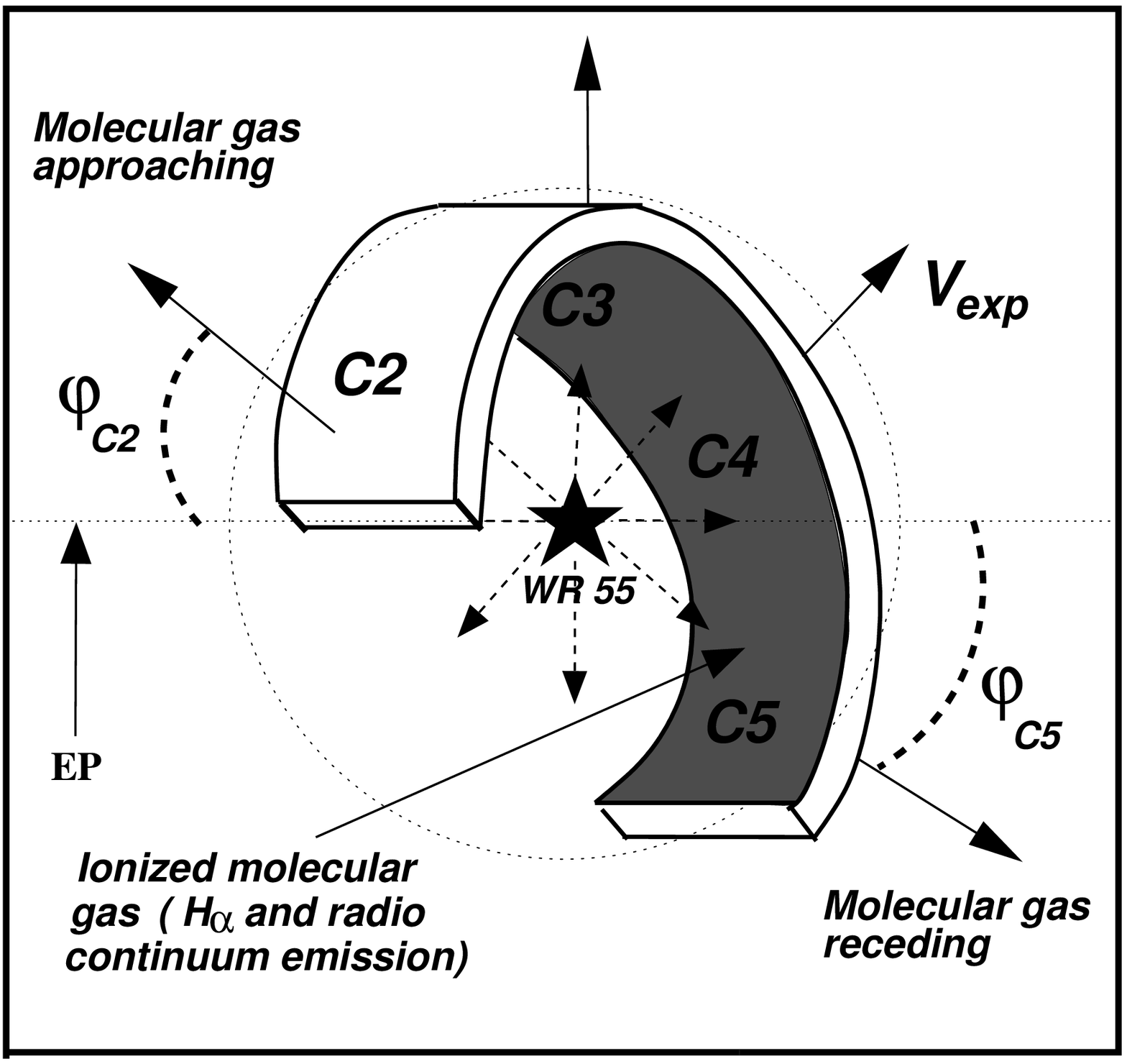}
\caption{Sketch of the model used for explaining the spatial distribution and velocity of MB.  The equatorial plane (EP) is depicted  by  the horizontal dotted line. 
}
\label{fig:rcw78modelo}
\end{figure}

Instead, concentrations C2 to C5 are best explained by an spherically expanding ring-like structure partially surrounding WR 55, and tilted with respect to the plane of the sky. A sketch of this model is given in Fig. {\ref{fig:rcw78modelo}}  where the approximate location of the different CO concentrations is given. The expansion of the molecular gas (filled arrows) is revealed by the velocity gradient (see Fig. \ref{fig:rcw78-bv}). The ultimate cause of the expanding molecular gas are the stellar winds (dashed line arrows) of WR 55. From here onwards the structure composed by concentrations C2, C3, C4, and C5 will be labelled molecular belt (or MB for short). To facilitate further analysis,  we divided MB into  two hemispheres with a plane which contains the WR star and the observer (equatorial plane). This plane (dubbed EP for short)  is also shown in Fig. \ref{fig:rcw78modelo}
  
\begin{figure}[h!]
\centering
\includegraphics[width=265pt]{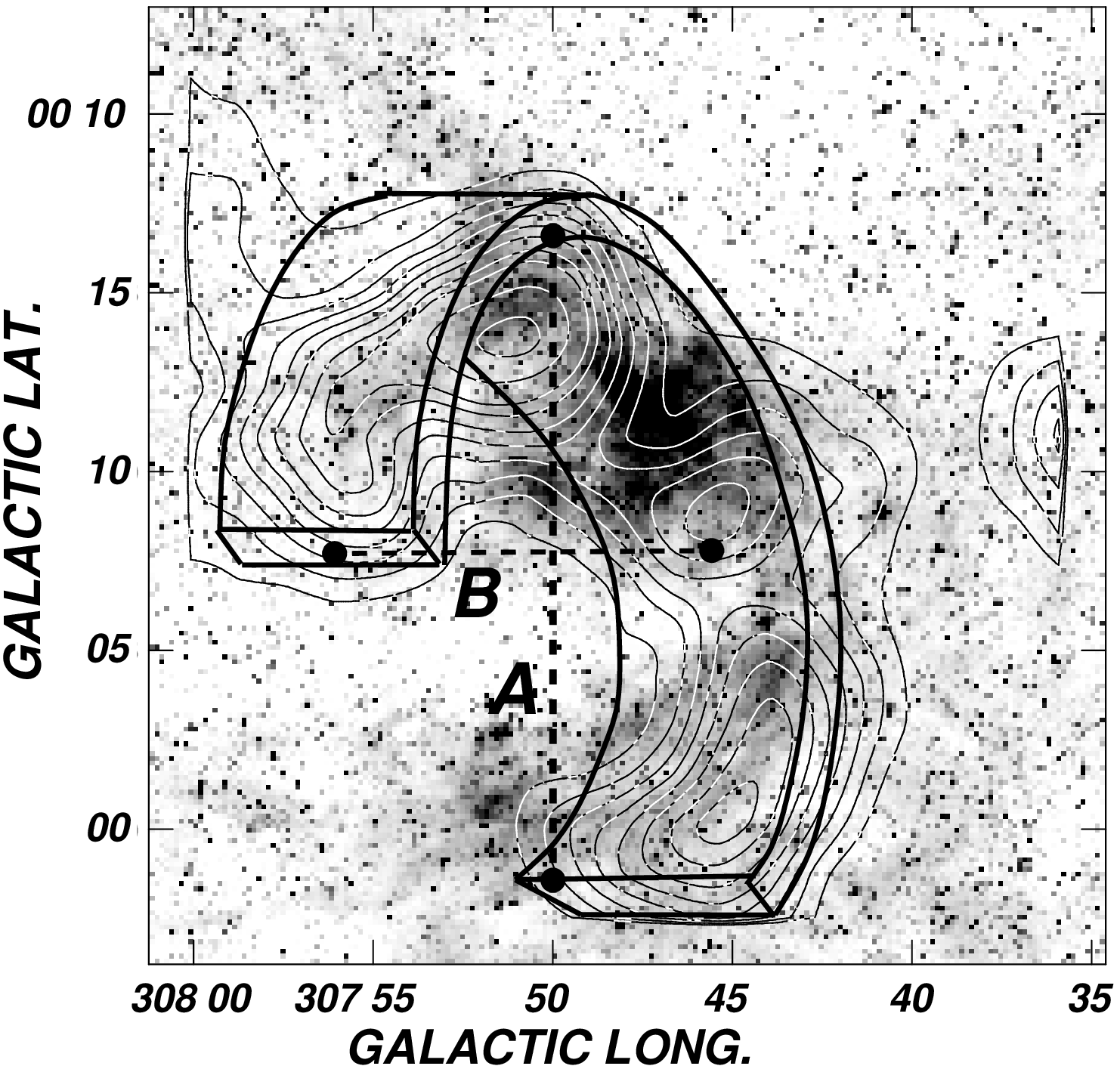}
\caption{Overlay of the model proposed for MB  (broad contours)   with  the molecular gas in the velocity range from $-$54 to $-$46 \kms   (narrow contours) around the nebula RCW 78. The major and minor semiaxis of MB (see text) are indicated by A and B, respectively.}
\label{fig:rcw78modelo-sup}
\end{figure}

Under the assumption that MB is expanding with respect to WR 55 with a velocity V$_{exp}$, the observed radial velocity is determined by two angles $\theta$ and $\varphi$. The former represents the inclination of MB with respect to the plane of the sky, whilst $\varphi$ represents the angle  between V$_{exp}$ and the plane EP  (see Fig. \ref{fig:rcw78modelo}). $\theta$ is single valued, whilst $\varphi$ can take any value between 0$^{\tiny \circ}$ and 360$^{\tiny \circ}$. For the sake of clarity only the $\varphi$ angles for concentrations C2 ($\varphi_{C2}$) and C5 ($\varphi_{C5}$) are shown in Fig. \ref{fig:rcw78modelo}. Based on the sketch shown in Fig. \ref{fig:rcw78modelo}, concentration C2 is approaching the observer (should show the most negative radial velocity) and concentration C5 is moving away from the observer (should show the most positive radial velocity). The other molecular concentrations (C3 and C4) should show intermediate radial velocities. A rough estimate of the angle $\theta$ can be obtained from the observed distribution of the molecular gas,  bearing in mind that a ring-like structure in space will become an elliptical feature when projected onto the  sky plane. In Fig. \ref{fig:rcw78modelo-sup}, the ring-like  distribution of our model is projected onto both the observed CO distribution in the velocity range from --54 to --46 \kms\ and the \ha\ image of RCW 78. The approximate minor/major semiaxis ratio (B/A)  of the elliptical structure can give an estimate of $\theta$ through the trigonometric relation B/A $\approx$ cos $\theta$. A value of B/A $\approx$ 0.5 can be   obtained, which leads to \hbox{$\theta$ $\sim$ 60$^{\circ}$}.

 Based on  Fig.  \ref{fig:rcw78modelo}, the maximum radial velocity difference among the molecular gas belonging to MB ($\Delta V_{\rm MB}$), will be given by the radial velocity difference between concentrations C2 ($V_{\rm  C2}$) and C5 ($V_{\rm  C5}$). From Table \ref{table:prop}, $\Delta V_{\rm MB}$ = $V_{\rm  C2}$ - $V_{\rm  C5}$ $\approx$ 6 \kms. Based in our model
\begin{eqnarray}
\qquad \Delta V_{\rm MB}\ =\ V_{exp}\ \times\ sin(\theta)\ \times\ \left[cos(\varphi_{C5})\ +\ cos(\varphi_{C2})\right]
\label{vel1} 
\end{eqnarray}
Using a coarse approximation $\varphi_{C2}$ $\approx$  $\varphi_{C5}$ (see Fig. \ref{fig:rcw78modelo}), Eq. \ref{vel1} can be written as
\begin{eqnarray}
\qquad \Delta V_{\rm MB}\ =\ 2\ \times\ V_{exp}\ \times\ sin(\theta)\ \times\ cos(\varphi_{C2})
\label{vel2} 
\end{eqnarray}
  Based on Fig. \ref{fig:rcw78modelo} it will be assumed that \hbox{$\varphi_{C2}$  $\approx$  45$^{\circ}$}. Admittedly, the uncertainty of $\varphi_{C2}$ can be very high.   Inserting the appropiate values of $\theta$ and $\varphi_{C2}$ in   Eq. \ref{vel2}, we obtain  $V_{\rm exp}$ $\approx$ 5  \kms. The main physical and geometrical properties  of MB are listed  in Table \ref{table:2}. The kinetic energy ($E_{\rm kin}$)  and momentum ($P$)  were obtained by considering an expansion velocity of $\sim$ 5 \kms.
 
The proposed model is able to explain  the main features of the optical nebula.  The inner face of MB, which is exposed to the far-UV radiation field of WR 55, may be ionized. This ionized layer gives rise to both the radio continuum and \Ha\ emission. Since concentration C2 is located between the observer and the ionized gas, it will diminish the intensity of the \ha\ emission that arises from the region of MB that faces WR 55. On the other hand, since towards concentrations C3, C4, and C5 the observer is directly viewing the inner ionized layer of MB, the \Ha\ emission of these regions is not absorbed by the molecular gas. Since concentrations C3 and C4 may be closer to WR 55 than C2 and C5, the intensity of both the \Ha\ and radio continuum emission is expected to be stronger there.

Though the proposed model is very simple, it is able to explain in a reasonable way the main observed characteristics of the \Ha, radio continuum, and molecular gas associated with RCW 78. Furthermore, the velocity gradient exhibited by the CO is also nicely accounted for.

 Certainly, more elaborated models are needed to better understand the origin and evolution of WRRN. Molecular ring-like structures  have been reported previously around several massive stars. Recently, \citet{bw10} surveyed the CO (3-2) line towards 43 identified {\it Spitzer} bubbles in the Galactic plane. They   concluded that the molecular gas tends to lie in rings, rather than shells. The authors  suggest that the parental molecular clouds, in which the massive stars and IBs are formed, are oblate with one dimension of a few parsecs in thickness. Then, expanding bubbles break out of the parental molecular cloud and only a circular or elliptical ring of CO emission is detected, depending on the orientation of the axis. The authors claimed that if the wind emitted by the star powering the bubble is sufficiently strong, the ring's expansion continues after breaching the flattened cloud. Nevertheless, \citet{d10} claimed that  a  possible shortcoming of the scenario proposed by \citet{bw10}   is the absence of a bipolar nebula as a result of a double {\it champagne flow} effect along the poles of the IB  \citep{tt79} .

\begin{table}
\caption{ Main geometrical and physical parameters of MB}
\label{table:2}
\begin{center}
\begin{tabular}{ l  r}
\hline\hline
 %&\\              
Parameter  &  Value\\
%&\\
\hline
%&\\
Major semiaxis ($'$)           & $\sim$ 9\\
%&\\
Minor semiaxis ($'$)           & $\sim$ 4.5\\
%&\\
Major semiaxis (pc)            & $\sim$  13\\
%&\\ 
$\Delta V_{\rm MB}$   (\kms)        & $\sim$ 6\\
%&\\        
$V_{\rm exp}$ (\kms)        &  $\sim$ 5 \\
%&\\
$V_{syst}$ (\kms)        &  $-$51\\
%&\\
$M_{\rm tot}$ (10$^4$ M$_{\odot}$)         &  3.4 $\pm$ 1.3\\
%&\\
$E_{\rm kin}$ ($10^{48}$ erg)  &  $\sim$ 9 \\
%&\\
 $P$  (10 $^4$ \msun \kms) &  $\sim$ 1.7  \\
\hline

\end{tabular}
\end{center}
\end{table}

\subsection{Energetics of RCW 78 and MB}

In our simple model, the main ionizing  source of RCW 78 is the Lyman continuum flux of the WR star WR 55. The latter is also the driving  source, via the mechanical energy injected by its stellar wind, of the expansion of the molecular gas associated with RCW 78. Is WR 55 capable of powering RCW 78 and providing the  mechanical energy needed to set in the expansion of the molecular gas?

To answer  the first part of the question, the number of Lyman  continuum photons needed to keep  the current level of ionization in RCW 78 must be calculated. To this end, the continuum flux densities of CF2, CF3, CF4, CF5, and CF6 were added up, and using  Eq. \ref{lyman} under the assumption that $T_e$ $\approx$ 10$^4$ K and $d$ $\approx$ 5 kpc, we derived  $N_{\nu}$ $\approx$ 1 $\times$ 10$^{48}$ s$^{-1}$. Considering that the number  of Lyman continuum photons emitted  by a WN7 star is $\sim$ 2.5 $\times$ 10$^{49}$ s$^{-1}$ \citep{cr07},  the WR star  may be capable of mantaining the ionization level of  RCW 78.

In regards to the point of whether the mechanical energy injected by WR 55 could be driving the expansion  of MB we shall derive a rough figure for the total mechanical energy injected via stellar winds by both WR 55 and its progenitor. The wind mechanical energy ($E_w$) will  be estimated using 
\begin{eqnarray}
 E_{\rm w}\ =\  L_{\rm w}\ t_{\rm w}\  =\ \frac{1}{2}\ \dot{M}\ V_{\rm w}^2\ t_{\rm d} 
\end{eqnarray}
where $\dot{M}$ is the  mass loss rate in units of \msunyr, $V_{\rm w}$ is  the wind terminal velocity in units of \kms, and   $t_{\rm d}$ is the time spent by the star during either the WR or the main sequence phase.  Adopting for WR 55 a mass loss rate $\dot{M}$ = 2$\times$10$^{-5}$ \hbox{M$_{\odot}$ yr$^{-1}$}, a terminal wind velocity $V_{\rm w}$ = 1960 \kms \citep{snc02}, and $t_{\rm WN}$ $\approx$ 5 $\times$ 10$^{5}$ yr \citep{vdh01}, a mechanical energy of $E_{\rm WN}$ $\approx$ 3.8 $\times$ 10$^{50}$ erg is obtained. Considering an O3V star as the progenitor of the WN star \citep{mass98}, and adopting $\dot{M}$ = 1.41 $\times$ 10$^{-6}$ \msunyr  and  V$_{\rm w}$ = 3150 \kms for this spectral type \citep{snc02}, and adopting a  main sequence lifetime of  $t_{\rm O}$ $\approx$ 2 $\times$ 10$^{6}$ yr \citep{mass98}, the mechanical energy injected  during the  main sequence is E$_{\rm O}$ $\approx$ 2.8 $\times$ 10$^{50}$ erg. Therefore, the total mechanical energy injected  into the interstellar medium is  $E_{\rm w_{tot}}$ $\approx$ 6.6 $\times$ 10$^{50}$ erg. Taking into account that the expanding kinetic energy of MB is  $E_{\rm kin}$ $\sim$ 9 $\times$ $10^{48}$ erg (see Table \ref{table:2}), only about  1.4$\%$  of the mechanical energy released by WR 55 and its projenitor would be needed to account for  the kinetic energy of the expanding molecular gas. This estimates is in good agreement with the conversion efficiencies  of the order of 2-5 $\%$ reported by H{\sc i} line studies of IBs \citep{cnhk96,cp01,cacps03}. Theoretical models  also suggest that in adiabatic wind bubbles  only a few percent of the injected mechanical energy will be converted into kinetic energy of the expanding gas   \citep{W77,k92,ar07}. Based on the above consideration, the stellar winds of both WR 55 and its progenitor are able to  provide the observed kinetic energy of MB.   In Regards to the momentum injected by the WR star and its progenitor to the ISM we obtain  $P_*$ $\approx$ 2.1 $\times$ 10$^4$ \msun \kms which is almost in agreement with the momentum of MB (see Table \ref{table:2}). This indicates that the momentum is better conserved in the system star-MB.

The dynamical age (yr) of a wind blown bubble in the momentum conserving case can be calculated using 
\begin{eqnarray}
t_{\rm d}\ =\ 0.5 \times 10^6\ \times\  \frac{R}{V_{\rm exp}}
\end{eqnarray}
\citep{mc83,lc99}, where $R$ is the radius of the bubble (pc), and $V_{\rm exp}$ is the expansion  velocity \hbox{(\kms)}. Adopting $R\approx$ 13 pc (linear length of major semiaxis) and $V_{\rm exp}$ = 5 \kms, a value of $t_{\rm d}$ $\approx$ 1.3 $\times$ 10$^6$ yr is obtained for MB. Considering uncertainties,   this value is almost in agreement with the duration of the O and WN phases combined.

 It is worth mentioning that   the small solid angle ($\ll$ 4$\pi$) subtended by MB reduces the wind  mechanical energy and ionizing  power  of WR 55  available to it. This may also help to explain the tightness in the energy requirements (radiative and mechanical) and dynamical age obtained before.

\section[]{Summary}

The $^{12}$CO (1-0)  and radio continuum emission distribution whithin a square region 25$'$ $\times$ 25$'$ in size centered on the WN7 star HD 117688 (WR 55) around the optical nebula RCW 78  have been analyzed using intermediate angular resolution CO data, high angular resolution radiocontinuum data at 1.4 GHz and 4.85 GHz, and \Ha\ data.

 The CO data  have allowed us to identify a molecular feature likely to be associated with RCW 78. This gas has a mean velocity of --49.5 \kms, and its spatial distribution shows an excellent morphological correlation with the optical nebula. This feature is far from being homogeneous and five molecular concentrations have been identified. Each one of them is very well correlated with different areas of RCW 78. These molecular concentrations exhibit a clear radial velocity gradient. The mean radial velocity of the different molecular concentrations and the radial velocity gradient shown by the CO is in good agreement with the \Ha\  data. The molecular gas related to RCW 78 has a total mass of  $\sim$ 3.4 $\times$  10$^4$ M$_{\odot}$

High resolution continuum data have revealed the presence of six  features towards  RCW 78. Five of them, identified as CF2, CF3, CF4, CF5, and CF6 are likely to be associated with the nebula. Based on radio continuum flux densitiy determinations at 1.4 and 4.85 GHz all these sources are thermal in nature. A sixth radio continuum fature (labelled CF1) is also thermal and very likely is related to HD 117797. Due to  the uncertainty in the distances of both HD 117797 and WR 55 (the WR star is associated with RCW 78) it is not clear  whether both objects are related to each other, or HD 117797 is a foreground object to RCW 78.

A very simple model has been elaborated in order to explain the radial velocity gradient depicted by the molecular gas, and same of the morphological properties shown by RCW 78. The model consits of an expanding ring-like structure of molecular gas, whose inner face is being ionized by the Lyman continuum photons emitted by WR 55. The ring-like feature is inclined by $\sim$ 60$^{\circ}$ with respect to the  sky plane, and has a low expansion velocity of $\sim$ 5 \kms. The WR star may well be the main ionization source of the nebula RCW 78 and the driving source of the expansion of the associated molecular gas.

\vspace{1cm}

\begin{acknowledgements}

 %We acknowledge the anonymous referees for their helpful comments that improved the presentation of this paper.
We acknowledge the referee, Prof. You-Hua Chu, and the editor, Malcolm Walmsley    for their helpful comments and suggestions that improved the presentation of this paper. This project was partially financed by the Consejo Nacional de Investigaciones Cient\'ificas y T\'ecnicas (CONICET) of Argentina under projects  \hbox{PIP 112-200801-01299}, Universidad Nacional de La Plata (UNLP) under project \hbox{11G/091}, and Agencia Nacional de Promoci\'on Cient\'ica y Tecnol\'ogica (ANPCYT) under project \hbox{PICT 14018/03}. 

This research has made use of the VIZIER database, operated at the CDS, Strasbourg, France.
We greatly appreciate the hospitality of all staff members of Las Campanas Observatory of the Carnegie Institute of Washington. We thank all members of the NANTEN staff, in particular Prof. Yasuo Fukui, Dr. Toshikazu Onishi, Dr. Akira Mizuno, and students Y. Moriguchi, H. Saito, and S Sakamoto. We also would like to thank Dr. D. Miniti (Pont\'{\i}fica Universidad Cat\'olica, Chile) and Mr. F Bareilles (IAR) for their involment in early stages of this project

\end{acknowledgements}

\bibliographystyle{aa}
\bibliography{bibliografia-wr55}
 
\IfFileExists{\jobname.bbl}{}
{\typeout{}
\typeout{****************************************************}
\typeout{****************************************************}
\typeout{** Please run "bibtex \jobname" to optain}
\typeout{** the bibliography and then re-run LaTeX}
\typeout{** twice to fix the references!}
\typeout{****************************************************}
\typeout{****************************************************}
\typeout{}

}

\end{document}